\documentclass[twocolumn,showpacs,preprintnumbers,
amsmath,amssymb,aps,prl]{revtex4}

\usepackage{bm}
\usepackage{graphicx}

\bibstyle{apsrev}

\begin{document}

\title{Absence of BCS Condensation of Atoms in a Repulsive Fermi Gas 
of Atoms}
\author{Fuxiang Han, Hailing Li, Minghao Lei, and E Wu}
\affiliation{Department of Physics,
Dalian University of Technology,
Dalian, Liaoning 116024,
China}
\date{\today}

\begin{abstract}
Possible Bose-Einstein condensation of molecules and 
Bardeen-Cooper-Schrieffer (BCS) condensation of Cooper pairs of atoms in 
an ultracold Fermi gas of atoms with a repulsive two-body interaction 
are studied by using the path integral representation of the grand 
partition function. From the self-consistent equations obtained in the 
present work for the order parameters, we have found that  BCS 
condensation of atoms can't occur in such a Fermi gas of atoms and that 
the condensate observed experimentally is composed of condensed 
molecules or possibly of preformed Cooper pairs. To substantiate our 
conclusions from the self-consistent equations for the order parameters, 
the effective atom-atom interaction mediated through molecules has also 
been computed with the Foldy-Wouthuysen transformation and has been 
found to be repulsive, which implies a net repulsive two-body 
interaction between atoms and hence the absence of BCS condensation of 
atoms.
\end{abstract}

\pacs{03.75.Ss, 05.30.Fk, 67.90.+z}


\maketitle

The experimental realization~\cite{DeMarco1999Science} of an ultracold 
Fermi gas of atoms marks the beginning of a new era of the scientific 
research on degenerate quantum Fermi gases. Since then, 
Bardeen-Cooper-Schrieffer (BCS) condensation of fermionic atoms and 
Bose-Einstein condensation (BEC) of diatomic molecules of these atoms as 
well as the crossover between the BCS and BEC regimes have been under 
intensive study both theoretically~\cite{Holland2001prl, 
Timmermans2001pla, Ohashi2002prl, Falco2004prl, Mackie2005prl, 
Baym2005np, Chen2005pr} and experimentally~\cite{Regal2003prl, 
Regal2004prl, Zwierlein2003prl, Zwierlein2004prl, Greiner2005prl, 
Partridge2005prl}, with the BCS-BEC crossover regime being easily 
reached with the magnetic-field Feshbach resonance~\cite{Regal2004prl}.

In the BCS-BEC crossover regime, the nature of condensation, 
\textit{i.e.}, whether the observed condensation is BCS condensation 
of Cooper pairs that are bound-states of fermionic atoms in momentum 
space or it is Bose-Einstein condensation of molecules that are 
bound-states in real space or it is condensation of preformed Cooper 
pairs of atoms, is still controversial~\cite{Falco2004prl}.  In the 
experimental study of the crossover regime, both the sign and the 
strength of the two-body interaction between atoms can be tuned with the 
magnetic-field Feshbach resonance.  It was reported that BCS 
condensation of atoms was observed on both the BCS (attractive 
two-body interaction) and BEC (positive two-body interaction) 
sides~\cite{Regal2004prl}.

In this Letter, we concentrate on the BEC side of the crossover regime 
and investigate the nature of condensation on this side. The Fermi gas 
of atoms is described by the generalized 
Hamiltonian~\cite{Holland2001prl}
\begin{eqnarray}
H &=& 2(\nu-\mu)\sum_{\bm{p}} a_{\bm{p}}^{\dagger}a_{\bm{p}}
+\sum_{\bm{k}\sigma}(\epsilon_{\bm{k}}-\mu)
c_{\bm{k}\sigma}^{\dagger}c_{\bm{k}\sigma}\nonumber\\
&&{}+\frac{U}{N}\sum_{\bm{k}\bm{k}^{\prime}\bm{p}}
c_{\bm{k}+\bm{p},\uparrow}^{\dagger}
c_{\bm{k}^{\prime}-\bm{p},\downarrow}^{\dagger}
c_{\bm{k}^{\prime}\downarrow}c_{\bm{k}\uparrow}\nonumber\\
&&{}+\frac{g}{\sqrt{N}}\sum_{\bm{p}\bm{k}}
\bigl(a_{\bm{p}}^{\dagger}c_{\bm{k}\uparrow}
c_{\bm{p}-\bm{k},\downarrow}+h.c.\bigr),
\label{Hamiltonian:operator form}
\end{eqnarray}
where $2(\nu-\mu)$ is the offset energy of a molecule relative to the 
chemical potential $\mu$, $a_{\bm{p}}^{\dagger}$ and $a_{\bm{p}}$ are 
creation and annihilation operators of a molecule of momentum $\bm{p}$ 
and energy $2(\nu-\mu)$, $c_{\bm{k}\sigma}^{\dagger}$ and 
$c_{\bm{k}\sigma}$ are creation and annihilation opperators of a 
fermionic atom of momentum $\bm{k}$, spin $\sigma$, and energy 
$\epsilon_{\bm{k}}=\hbar^2k^2/2m$, $U$ ($>0$) is the strength of the 
two-body interaction between atoms, $g$ is the effective atom-molecule 
coupling constant, and $N$ is the average number of atoms. Here the spin 
up and spin down actually represent two different hyperfine levels of 
atoms. Here we have assumed that there are only two components in the 
gas.

Notice that the number of atoms $N$ in the system appears in the 
prefactors of the two-body interaction and atom-molecule coupling terms 
due to the Fourier transformation of the operators into momentum space. 
The presence of $N$ in these prefactors in conjunction with the proper 
change of variables to be performed below will give the self-consistent 
equations for the order parameters the proper system size dependence.

The aim of the present work is to study the nature of condensation on 
the BEC side of the crossover regime with the Hamiltonian given in 
Eq.~\eqref{Hamiltonian:operator form}. The nature of condensation is 
described by the corresponding order parameter. Therefore, to study 
the nature of condensation, we first set up the self-consistent 
equations for the order parameters of all possible types of 
condensation. Here we are concerned with two types of condensation, 
Bose-Einstein condensation of molecules and BCS condensation of atoms. 
The order parameter for Bose-Einstein condensation of molecules is the 
number of molecules in the lowest-energy level, and the order parameter 
for BCS condensation of atoms is the energy gap in the excitation 
spectrum of quasiparticles. Our criterion is that a certain type of 
condensation is said to occur if the corresponding order parameter has a 
nontrivial solution below a certain temperature.

Our starting point is the path integral representation~\cite{Negele1998, 
Zinn-Justin1996} of the grand partition function 
$Z=\mathrm{Tr}\,e^{-\beta H}$. The order parameter for Bose-Einstein 
condensation is introduced in the path integral representation of the 
grand partition function, whereas the order parameter for BCS 
condensation of atoms is introduced through the Hubbard-Stratonovich 
transformation~\cite{Hubbard1972pl, Stratonovich1958spd}. The 
path-integral representation of the grand partition function 
corresponding to the Hamiltonian in 
Eq.~\eqref{Hamiltonian:operator form} is given by~\cite{Negele1998, 
Zinn-Justin1996}
\begin{eqnarray}
Z &=& \int D[\varphi^{\ast},\varphi;\psi^{\ast},\psi]_{\tau}\;
\exp\left\{-\int_0^{\beta} 
d\tau\sum_{\bm{p}}\biggl[
\varphi_{\bm{p}}^{\ast}(\tau)\right.\nonumber\\
&&{}\times\bigl[\partial/\partial\tau 
+2(\nu-\mu)\bigr]\varphi_{\bm{p}}(\tau)\nonumber\\&&{}
+\sum_{\bm{k}\sigma}
\psi_{\bm{k}\sigma}^{\ast}(\tau)\bigl[\partial/\partial\tau
+(\epsilon_{\bm{k}}-\mu)\bigr]\psi_{\bm{k}\sigma}
(\tau)\nonumber\\
&&{}+\frac{U}{N}\sum_{\bm{k}\bm{k}^{\prime}\bm{p}}
\psi_{\bm{k}+\bm{p},\uparrow}^{\ast}(\tau)
\psi_{\bm{k}^{\prime}-\bm{p},\downarrow}^{\ast}(\tau)
\psi_{\bm{k}^{\prime}\downarrow}(\tau)
\psi_{\bm{k}\uparrow}(\tau)\nonumber\\
&&{}\left.+\frac{g}{\sqrt{N}}\sum_{\bm{p}\bm{k}}
\bigl[\varphi_{\bm{p}}^{\ast}(\tau)
\psi_{\bm{k}\uparrow}(\tau)
\psi_{\bm{p}-\bm{k},\downarrow}(\tau)
+c.c.\bigr]\biggr]\right\},\;\quad
\label{partition function:image time}
\end{eqnarray}
where $c.c.$ denotes the complex conjugate of the proceeding term, 
$\varphi_{\bm{p}}^{\ast}(\tau)$ and $\varphi_{\bm{p}}(\tau)$ are complex 
variables corresponding to the creation and annihilation operators 
$a_{\bm{p}}^{\dagger}(\tau)$ and $a_{\bm{p}}(\tau)$ of a bosonic 
molecule, and $\psi_{\bm{k}\sigma}^{\ast}(\tau)$ and 
$\psi_{\bm{k}\sigma}(\tau)$ are Grassman variables corresponding to the 
creation and annihilation operators $c_{\bm{k}\sigma}^{\dagger}(\tau)$ 
and $c_{\bm{k}\sigma}(\tau)$ of a fermionic atom. The integration 
measure $D[\varphi^{\ast},\varphi,\psi^{\ast},\psi]_{\tau}$ is given by
\begin{eqnarray}
D[\varphi^{\ast},\varphi;\psi^{\ast},\psi]_{\tau}
&=& \lim_{M\rightarrow\infty}\prod_{j=1}^{M}\prod_{\bm{k}\sigma}
d\psi_{\bm{k}\sigma}(j\Delta\tau)
d\psi_{\bm{k}\sigma}^{\ast}(j\Delta\tau)\nonumber\\
&&{}\times
\prod_{\bm{p}}\frac{d\varphi_{\bm{p}}^{\ast}(j\Delta\tau)
d\varphi_{\bm{p}}(j\Delta\tau)}{2\pi i}
\end{eqnarray}
with $\Delta\tau=\beta/M$.

The path-integral representation of the grand partition function in 
Eq.~\eqref{partition function:image time} is in imaginary time space and 
it can be transformed into imaginary frequency space through Fourier 
transformations of the complex and Grassman variables
\begin{subequations}
\begin{eqnarray}
&& \varphi_{\bm{p}}(\tau) 
= \sum_{i\omega_m}e^{-i\omega_{m}\tau}\varphi_{\bm{p}}(i\omega_m),\\
&& \psi_{\bm{k}\sigma}(\tau) = \sum_{i\omega_n}
e^{-i\omega_{n}\tau}\psi_{\bm{k}\sigma}(i\omega_n),
\end{eqnarray}
\end{subequations}
where $i\omega_m=i2m\pi/\beta$ and $i\omega_n=i(2n+1)\pi/\beta$,
with $m$ and $n$ being integers, are Matsubara imaginary frequencies for 
bosons and fermions, respectively. After the Fourier transformations, 
the complex and Grassman variables are denoted succinctly by 
$\varphi_{p}$ and $\psi_{k\sigma}$, respectively, with short notations 
$p=(\bm{p}, i\omega_m)$ and $k=(\bm{k}, i\omega_n)$ introduced. The BCS 
order parameter $\Delta_{p}$ is introduced through the 
Hubbard-Stratonovich transformation performed with respect to the 
two-body interaction term that, for convenience of the performance of 
this transformation, is recast into the following form
\begin{eqnarray}
\lefteqn{\frac{U}{N}\sum_{kk^{\prime}p}\psi_{k+p,\uparrow}^{\ast}
\psi_{k^{\prime}-p,\downarrow}^{\ast}\psi_{k^{\prime}\downarrow}
\psi_{k\uparrow}}\nonumber\\
&=& U\sum_{p}\biggl[\frac{1}{\sqrt{N}}\sum_{k}
\psi_{p-k\downarrow}\psi_{k\uparrow}\biggr]^{\ast}
\biggl[\frac{1}{\sqrt{N}}\sum_{k}
\psi_{p-k\downarrow}\psi_{k\uparrow}\biggr].\;\quad
\end{eqnarray}

For the repulsive two-body interaction term, the Hubbard-Stratonovich 
transformation must be implemented by making use of the Gaussian 
integral of the form $e^{-\zeta^{\ast}\zeta/\alpha} = (\alpha/2\pi i) 
\int dz^{\ast}dz\;e^{-\alpha z^{\ast}z + \zeta z^{\ast} - 
\zeta^{\ast}z}$ with $z$ and $\zeta$ complex variables. That the two 
decoupled terms have different signs is because the two-body term has a 
minus sign on the exponential in Eq.~\eqref{partition function:image 
time}. If an attractive two-body interaction term were to be decoupled, 
the two decoupled terms would have identical signs. Here the auxiliary 
field introduced in the Hubbard-Stratonovich transformation is denoted 
by $\Delta_{p}$ which is the order parameter for BCS condensation of 
atoms. It has been found that, to derive self-consistent equations for 
$\varphi_{p}$ and $\Delta_{p}$ with a proper system size dependence, it 
is necessary to make a change of integration variables $\varphi_{p} 
\rightarrow \sqrt{N} \varphi_{p}$ and $\Delta_{p} \rightarrow \sqrt{N} 
\beta U \Delta_{p}$.

For the purpose of studying Bose-Einstein condensation of molecules 
and BCS condensation of atoms, it is sufficient to retain only the 
zero momentum and frequency terms in the atom-molecule coupling term and 
in the decoupled terms arising from the two-body interaction. Then, the 
integration over the order parameters $\varphi_{p}$ and $\Delta_{p}$ 
with $p\neq 0$ can be performed and the summation over $i\omega_m$ can 
be completed afterwards. And then the Grassman variables can be 
integrated out.

To integrate out the Grassman variables, we make a change of variables 
to bring the relevant terms into a diagonal form, with the process 
similar to the diagonalization of the Hamiltonian of a system of 
fermions. For this purpose, we reexpress the terms containing the 
Grassman variables into a matrix form through an introduction of a 
column matrix of Grassman variables. The resulting $2\times 2$ matrix is 
to be diagonalized. Since this  $2\times 2$ matrix is not Hermitian,  a 
proper care must be given to the finding of its eigenvalues and  
corresponding right eigenvectors. With the Grassman variables integrated 
out, the summation over the imaginary frequencies $i\omega_n$ can be 
performed.

Lastly, we consider only the real parts of the order parameters. For the 
real order parameters, the imaginary part of the form $i\mathrm{Im}\, 
(\varphi^{\ast} \Delta - \Delta^{\ast} \varphi)$ in the contribution to 
the action arising from the integration of Grassman variables is 
identically zero. With this imaginary part identified to be zero, the 
imaginary parts of the order parameters in the other contributions to 
the action can be integrated out, which provides a reasonable procedure 
for retaining only the real parts of the order parameters.

Finally, neglecting all the irrelevant prefactors in the partition 
function arising from the above-described algebraic manipulations, we 
obtain the following path-integral representation of the grand partition 
function
\begin{equation}
Z \sim \int d\varphi d\Delta\; e^{-S},
\end{equation}
where $S$ is referred to as the action and is given by
\begin{eqnarray}
S &=& \sum_{\bm{p}\neq 0}
\ln\frac{\sinh(\beta\Delta\nu)}{\beta\Delta\nu}
-2\sum_{\bm{k}}\ln\biggl(2
\cosh\frac{\beta E_{\bm{k}}}{2}\biggr)\nonumber\\
&&{} + 2N\beta \Delta\nu \varphi^{2}
+N\beta U\Delta^{2}
\label{action:final}
\end{eqnarray}
with $\Delta\nu=\nu-\mu$ and
\begin{equation}
E_{\bm{k}} = (\xi_{\bm{k}}^2+g^2\varphi^{2} - U^2 
\Delta^{2})^{1/2}.
\end{equation}
Here $\varphi$ and $\Delta$ are the order parameters of zero momentum 
and frequency. Notice that the summation over $\bm{p}$ in the first term 
on the right hand side of Eq.~\eqref{action:final} yields a factor 
proportional to the number of molecules with $\bm{p}\neq 0$.

We now infer information about the order parameters for Bose-Einstein 
condensation of molecules and BCS condensation of atoms from the above 
path integral representation of the grand partition function. The 
integrals in the grand partition function can be performed in principle 
by using the steepest descent method. The stationary point in the 
order parameter space in this method corresponds to the 
Hartree-Fock-Bogoliubov approximation to the order parameters. To obtain 
an effective action, we proceed as follows. With the action on the 
exponential being expanded as a Taylor series of the order parameters 
about the stationary point and with the terms up to the second order 
being kept, the integration in Eq.~\eqref{action:final} can be 
performed. The effective action in the Hartree-Fock-Bogoliubov 
approximation is actually obtained by retaining only the 
stationary-point contribution to the action.

Taking partial derivatives of the action with respect to $\varphi$ and 
$\Delta$ and setting them to zero to find the stationary values of 
$\varphi$ and $\Delta$, we obtain the self-consistent equations for 
$\varphi$ and $\Delta$
\begin{subequations}
\label{sces}
\begin{eqnarray}
&& \varphi
= \varphi\frac{g^2}{4N\Delta\nu}\sum_{\bm{k}}
\frac{\tanh(\beta E_{\bm{k}}/2)}{E_{\bm{k}}},
\label{sce:varphi}\\ 
&& \Delta = -\Delta\frac{U}{2N} \sum_{\bm{k}}
\frac{\tanh(\beta E_{\bm{k}}/2)}{E_{\bm{k}}}.
\label{sce:Delta}
\end{eqnarray}
\end{subequations}

The self-consistent equations of the order parameters $\varphi$ and 
$\Delta$ in Eqs.~\eqref{sces} are the central results derived in the 
present work and they provide the basis for our following discussions 
and conclusions. Noticing that $U>0$ for the Fermi gas of atoms under 
our current study and the value of the wavevector sum in 
Eq.~\eqref{sce:Delta} is positive, we immediately realize that there 
exists only a trivial solution for $\Delta$ from Eq.~\eqref{sce:Delta}, 
which implies that fermionic atoms can't condense via the mechanism of 
BCS condensation! However, the self-consistent equation for $\varphi$ in 
Eq.~\eqref{sce:varphi} may have a nontrivial solution. We can thus 
conclude that possible condensation or superfluidity in a repulsive 
Fermi gas of atoms is that of molecules or preformed Cooper pairs.

Algebraically, the nonexistence of a nontrivial solution for $\Delta$ is 
because there is a minus sign in front of $\Delta^2$ in the excitation 
spectrum of quasiparticles as can be seen from $E_{\bm{k}} = 
(\xi_{\bm{k}}^2 + g^2 \varphi^2 - U^2 \Delta^2)^{1/2}$. This minus sign 
leads to a minus sign on the right hand side of the self-consistent 
equation for $\Delta$.

Physically, the nonexistence of a nontrivial solution for $\Delta$ is 
because of the absence of an effective \textit{attractive} two-body 
interaction among atoms. From the Hamiltonian in 
Eq.~\eqref{Hamiltonian:operator form}, it is seen that the only possible 
physical process that might generate an attractive two-body interaction 
among atoms is the atom-molecule coupling. However, the effect of this 
coupling is actually pair breaking in that a pair is lift out of its 
condensed state and transformed into a molecule. The reverse process 
does not lead to pairing in that it can only be said that a molecule is 
dissolved into two atoms that occupy two states with opposite momenta. 
However, for pairing to occur, there must exist an effective attractive 
interaction between these two atoms.

We now investigate whether an effective attractive interaction can be 
produced by the atom-molecule coupling. In analogy to the 
phonon-mediated electron-electron interaction in metals, it can be 
explicitly shown that the effective interaction between atoms mediated 
through molecules is repulsive. For this purpose, we consider the 
following Hamiltonian $H=H_{0}+H_{1}$ with
\begin{subequations}
\begin{eqnarray}
&& H_{0} = 2\Delta\nu a^{\dagger}a
+ \sum_{\bm{k}\sigma}\xi_{\bm{k}}
c_{\bm{k}\sigma}^{\dagger}c_{\bm{k}\sigma},\\
&& H_{1} = \frac{g}{\sqrt{N}}\sum_{\bm{k}}\bigl(
a^{\dagger}c_{\bm{k}\uparrow}c_{-\bm{k}\downarrow}+h.c.\bigr).
\end{eqnarray}
\end{subequations}
Here for simplicity we included in the Hamiltonian only the 
contributions from zero-momentum molecules and we shall consider the 
effective atom-atom interaction mediated through these molecules. The 
effective atom-atom interaction mediated through nonzero-momentum 
molecules is qualitatively similar.

To infer the effective atom-atom interaction mediated through molecules, 
we adopt the procedure due to Foldy and Wouthuysen~\cite{Foldy1950pr, 
Bjorken1964} and make a canonical transformation to the above 
Hamiltonian
\begin{equation} 
H^{\prime} = e^{-S} H e^{S}
\end{equation}
with $S$ assumed to be of the form
\begin{equation} 
S = \sum_{\bm{k}}\bigl( A_{\bm{k}} a^{\dagger} c_{\bm{k}\uparrow} 
c_{-\bm{k}} + B_{\bm{k}} a c_{-\bm{k}}^{\dagger} 
c_{\bm{k}\uparrow}^{\dagger} \bigr).
\end{equation} 
Expanding $H^{\prime}$ and keeping terms up to the second order in $S$, 
we have $H^{\prime} \approx H + [H, S] + \frac{1}{2}[[H_{0}, S], S]$. 
The coefficients in $S$ are to be determined from the condition that 
$H_{1} + [H_{0}, S] = 0$. The transformed Hamiltonian is then 
approximately given by
\begin{equation}
H^{\prime} \approx H_{0} + \frac{1} {2}[H_{1}, S].
\end{equation}
The concerned effective atom-atom interaction mediated through molecules 
is to be deduced from the second term in the above equation. With only 
the atom-atom two-body terms retained, the effective atom-atom 
interaction is given by
\begin{equation} 
\frac{g^2}{2}\sum_{\bm{k}\bm{k}^{\prime}}
\Bigl(\frac{1}{\Delta\nu-\xi_{\bm{k}}}
+\frac{1}{\Delta\nu-\xi_{\bm{k}}^{\prime}}\Bigr)
c_{-\bm{k}\downarrow}^{\dagger}c_{\bm{k}\uparrow}^{\dagger}
c_{\bm{k}^{\prime}\uparrow}c_{-\bm{k}^{\prime}\downarrow}.
\end{equation}

From the above equation, it is  clearly seen that the effective 
atom-atom interaction mediated through molecules is repulsive for atoms 
close to the Fermi surface as long as the molecule offset energy  
$\Delta\nu = \nu - \mu$ is greater than zero. Since atoms close to the 
Fermi surface play a dominant role in BCS condensation and since 
$\Delta\nu>0$ is satisfied in the concerned Fermi gas of 
atoms~\cite{Regal2004prl}, we thus conclude that the effective atom-atom 
interaction mediated through molecules is repulsive.

For atoms in a repulsive Fermi gas of atoms used in the recent
experiments~\cite{Regal2004prl}, in addition to the repulsive two-body 
interaction tunable through the Feshbach resonance, there exists also 
the above-calculated effective interaction mediated through molecules. 
According to BCS theory~\cite{Cooper1956pr, Bardeen1957pr}, the 
formation and immediate condensation of Cooper pairs of fermionic atoms 
require that the net two-body interaction between atoms be attractive. 
From the fact that both the original two-body interaction and the 
effective two-body interaction mediated through molecules are repulsive, 
it follows that the net two-body interaction is repulsive, which implies 
the impossibility of BCS condensation of atoms, in consistency with 
the conclusions drawn previously from the self-consistent equations for 
the order parameters. Therefore, the observed condensation on the BEC 
side can't be BCS condensation of atoms. It may be condensation of 
molecules or that of preformed Cooper pairs. Notice that condensation of 
preformed Cooper pairs distingushes from BCS condensation of atoms in 
that the former is a two-step process in which Cooper pairs first come 
into being as bound states of atoms and then condense, while the latter 
is a single-step process in which the formation and condensation of 
Cooper pairs occur simultaneously. If condensation on the BEC side 
observed in the experiment turns out to be that of preformed Cooper 
pairs, then the mechanism of the formation of these Cooper pairs will 
certainly shed light on the mechanism of high temperature 
superconductivity.

To summarize, within the path-integral formalism of the grand partition 
function, we have derived self-consistent equations for the order 
parameters of Bose-Einstein condensation of molecules and BCS 
condensation of atoms in a Fermi gas of atoms with a repulsive two-body 
interaction. We have found that there is only trivial solution of the 
order parameter for BCS condensation of atoms, from which we conclude 
that BCS condensation of atoms is not achievable in such a Fermi gas of 
atoms and that the observed condensation must be that of molecules or 
that of preformed Cooper pairs. The explicit calculation of the 
effective atom-atom two-body interaction mediated through molecules 
demonstrates that there is no net attractive interaction between atoms 
and therefore confirms our conclusions.

\bibliography{repulsive}

\end{document}